# A Monte Carlo assessment of the spectral performance of four types of photon counting detectors

Karl Stierstorfer, Martin Hupfer
Siemens Healthineers, Siemensstr. 3, Forchheim, Germany

**ABSTRACT**

In previous publications, we have presented an alternative approach to determine essential detector properties like the Modulation Transfer Function (MTF), the Noise Power Spectrum (NPS) and the Detective Quantum Efficiency (DQE) based on a Monte Carlo model of the detection process. If a Monte Carlo model for the detector response to photons impinging at various locations of a pixel is available, the full statistics of the detector can be derived in a straightforward manner.

The purpose of this paper is to describe the method in detail and to apply it to four types of realistic detectors: direct converting detectors using CdTe and silicon, a CdTe photon counter with additional coincidence counters and an optical counting system using $LaBr_3$ as scintillator.

## 1. INTRODUCTION

Assessing the statistical performance of detectors is crucial for the design of new or the optimization of existing systems. Information transmission in detectors can be quite complex, including various interconnected mechanisms like amplification/reduction or scatter/crosstalk. The standard approach (cf. e.g. [1]) is to decompose the detection process into a cascade of elementary processes with simple statistics. While this approach may provide additional insights by making the contributions of each elementary process transparent in analytic equations, it is often cumbersome to perform, and the final expressions can become unwieldy.

Based on the elegant technique of moment-generating functions, Rabbani et al. [2] developed a framework for the calculation of MTF and DQE of detectors including cascaded amplification and scatter steps. To perform the calculations on a pixelized grid, a virtual pixel size $\Delta$ was introduced. Since the framework was used for analog (e.g., screen-film) detector systems, a $\lim_{\Delta \to 0}(\cdot)$ was performed at the end of the calculation. In [3], we used the same method of moment-generating functions with the aim to calculate the statistics of photon counting detectors but, since we were applying it for truly pixelized detectors, there was neither need nor justification to perform a limit towards infinitely small pixel size. Instead, to correctly describe the response of the system to input signals in the *analog* input domain, as is crucial for the calculation of the *pre-sampling* MTF, we introduced the *impact location* **u**, the location of a photon hitting the detector relative to the pixel center.

The goal of this paper is 1) to explain in detail how the method can be used also to assess the *spectral* performance of detectors with several spectral channels and 2) to apply the method to four different spectral detectors: a) a CdTe based direct converter, b) a Si base direct converting detector [9], c) a CdTe based direct converter with additional coincidence counters [7], d) an optical counting detector based on $LaBr_3$ as scintillator. The two last systems have not yet been realized.

## 2. MATERIAL AND METHODS

### 2.1 Basic equations

Performing analytic calculations for a general linear detector in the low-count limit, we derived in [3] simple equations relating both mean values and covariances for the output random variables to the statistics of the input random variable and the statistics of the detection process. Using bold characters $\mathbf{m} = (m_1, m_2)$, $\mathbf{u} = (u_1, u_2)$ etc. to denote the 2D vectors characterizing the 2D geometry of the detector (see figure 1), we denote the output random variable of pixel signals by $y_\mathbf{m}$ and the input random variable by $x_\mathbf{m}(\mathbf{u})$. The link between the input and the output is conveyed by the random variable $a_\mathbf{m}^{(\mathbf{k})}(\mathbf{u})$ describing the response in pixel $\mathbf{m}$, given a single photon is injected in pixel $\mathbf{k}$ at impact location $\mathbf{u}$ relative to the pixel center.

(1) $\overline{y_\mathbf{m}}(\mathbf{u}) = \sum_\mathbf{k} \overline{a_\mathbf{m}^{(\mathbf{k})}(\mathbf{u})} \, \overline{x_\mathbf{k}}(\mathbf{u}) = \sum_\mathbf{k} \overline{a_{\mathbf{m}-\mathbf{k}}^{(0)}(\mathbf{u})} \, \overline{x_\mathbf{k}}(\mathbf{u}),$

(2) $\mathrm{cov}_\mathbf{u}(y_\mathbf{m}, y_\mathbf{n}) = \sum_{\mathbf{kl}} \overline{a_\mathbf{m}^{(\mathbf{k})}(\mathbf{u}) a_\mathbf{n}^{(\mathbf{l})}(\mathbf{u})} \, \mathrm{cov}_\mathbf{u}(x_\mathbf{k}, x_\mathbf{l}) + \sum_\mathbf{k} \mathrm{cov}_\mathbf{u}\left(a_\mathbf{m}^{(\mathbf{k})}, a_\mathbf{n}^{(\mathbf{k})}\right) \overline{x_\mathbf{k}} =$

$N \sum_\mathbf{k} \overline{a_\mathbf{m}^{(\mathbf{k})} a_\mathbf{n}^{(\mathbf{k})}}(\mathbf{u}) = N \sum_\mathbf{k} \overline{a_{\mathbf{m}-\mathbf{k}}^{(0)} a_{\mathbf{n}-\mathbf{k}}^{(0)}}(\mathbf{u}),$

where the last steps assume Poisson statistics and homogeneous illumination for the input $x_\mathbf{m}$: $\mathrm{cov}(x_\mathbf{k}, x_\mathbf{l}) = \overline{x_\mathbf{k}} \delta_{\mathbf{kl}} = N \delta_{\mathbf{kl}}$ and a uniform (translationally symmetric) detector: $a_\mathbf{m}^{(\mathbf{k})} = a_{\mathbf{m}-\mathbf{k}}^{(0)}$. The bar sign expresses averaging across photon histories.

Given the statistics of the random variable $a_\mathbf{m}^{(0)}(\mathbf{u})$ for the signal (counts) generated in pixel $\mathbf{m}$ from photons impinging on pixel (0,0) at impact location $\mathbf{u}$, the first equation allows calculating the pre-sampling modulation transfer function (MTF). Likewise, the autocorrelation function and thus the noise power spectrum (NPS) can be derived by averaging the expression $\sum_\mathbf{k} \overline{a_{\mathbf{m}-\mathbf{k}}^{(0)} a_{\mathbf{n}-\mathbf{k}}^{(0)}}(\mathbf{u})$ across all impact locations $\mathbf{u}$. For details, refer to [3].

Note that no point in the derivation of the above equations has required any assumptions regarding the statistics of the *output* variables $y$. Thus, $y$ could also be the number of electrons created in the photo diodes of a detector using scintillators. Thus, it would be straightforward to use the formalism to describe the statistics of energy integrating detectors.

In the following, we are considering detectors with several spectral channels, for example several thresholds or bins. In the above formalism, they are taken into account by giving the random variables $y$ and $a$ an additional superscript. Also, we need to consider that all quantities will depend on the energy of the incoming photon $\epsilon$:

(3) $\overline{y_\mathbf{m}^s}(\mathbf{u}, \epsilon) = \sum_\mathbf{k} \overline{a_\mathbf{m}^{(\mathbf{k}),s}(\mathbf{u}, \epsilon)} \, \overline{x_\mathbf{k}}(\mathbf{u}, \epsilon) = \sum_\mathbf{k} \overline{a_{\mathbf{m}-\mathbf{k}}^{(0),s}(\mathbf{u}, \epsilon)} \, \overline{x_\mathbf{k}}(\mathbf{u}, \epsilon),$

the covariance becomes a 2-D matrix in the spectral indices:

(4) $\mathrm{cov}_{\mathbf{u},\epsilon}(y_\mathbf{m}^s, y_\mathbf{n}^{s'}) = N(\epsilon) \sum_\mathbf{k} \overline{a_{\mathbf{m}-\mathbf{k}}^{(0),s}(\mathbf{u}, \epsilon) a_{\mathbf{n}-\mathbf{k}}^{(0),s'}(\mathbf{u}, \epsilon)}.$

## 2.2 How to obtain the spectral Modulation Transfer Function and the spectral Noise Power Spectrum from a Monte Carlo simulation

The procedure to obtain the MTF and NPS directly from a Monte Carlo simulation is now presented in detail. We assume the center of pixel $\mathbf{0} = (0,0)$ to be located at the center of the coordinate system at $\mathbf{u} = (0,0)$ (see Figure 1). $\Delta$ is the pixel size. For simplicity, we are considering quadratic pixels here, but a generalization to detectors with non-square pixels would be straightforward.

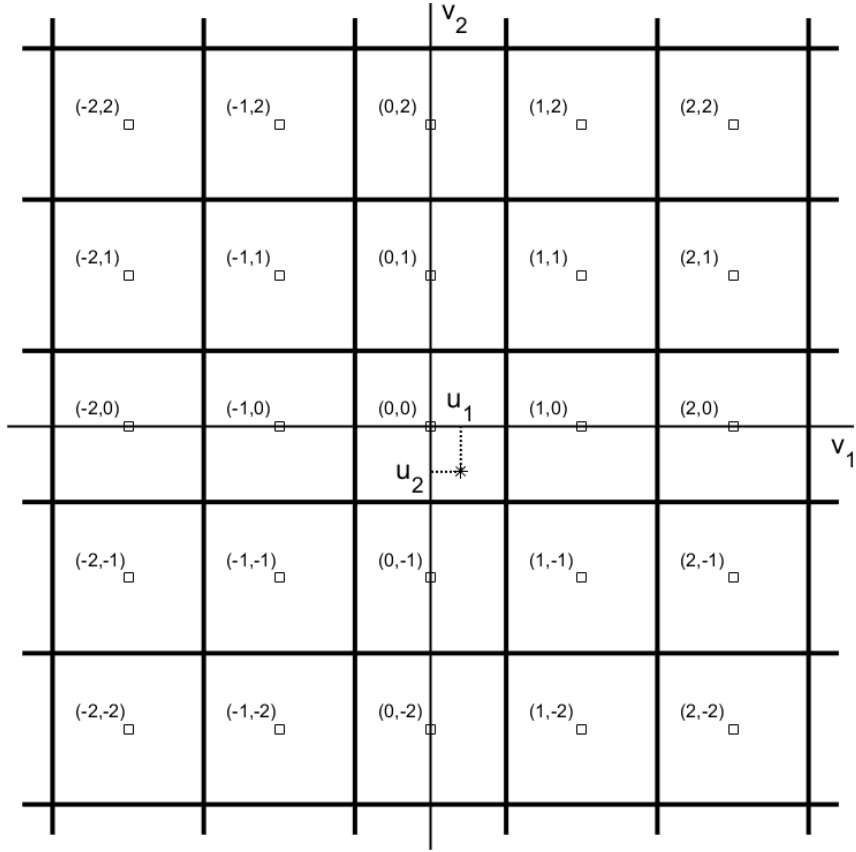

*Figure 1: Basic geometry showing a 5x5 section of the pixel array. The squares indicate the centers of the pixels. The photon is hitting the detector in pixel (0,0) at impact location $\mathbf{u} = (u_1, u_2) = (0.2, -0.3)\Delta$ (marked with a star).*

These are the steps to obtain the spectral MTF and NPS, starting with a Monte Carlo simulation of the detection process:

- For each photon energy $\epsilon$ and each $\mathbf{u} = (u_1, u_2) \in [-0.5\Delta, 0.5\Delta] \times [-0.5\Delta, 0.5\Delta]$ on some reasonable grid, e.g. 100x100, perform $P$ Monte Carlo simulations of the detector response from a single photon hitting the central pixel at impact location $\mathbf{u}$, providing the $P$ vectors $\gamma^s_{\mathbf{m},p}(\mathbf{u}, \epsilon), p = 1 \ldots P$ of signals in all pixels $\mathbf{m} = (m_1, m_2)$ and spectral channels $s$. Typically, $P = 1000$ is a reasonable figure. Note that, for photon counting detectors, the $\gamma^s_{\mathbf{m},p}(\mathbf{u}, \epsilon)$ are either 0 or 1. Double (or multiple) counting arises from the fact that an incoming photon may create counts in *several* pixels $\mathbf{m}$.
- Calculate the pre-sampling point spread function:

(5) $\quad \text{PSF}^s(\mathbf{v} = \mathbf{m}\Delta - \mathbf{u}, \epsilon) = \frac{1}{P}\sum_{p=1}^{P} \gamma^s_{\mathbf{m},p}(\mathbf{u}, \epsilon).$

- Calculate the covariance matrix by averaging products across the incoming photons $p$ and impact locations $\mathbf{u}$:

(6) $\quad \mathrm{cov}_\epsilon(y_\mathbf{m}^S, y_\mathbf{n}^{S'}) = \frac{1}{\Delta^2} \iint_{-\Delta/2}^{\Delta/2} \sum_p \gamma_{\mathbf{m},p}^S(\mathbf{u}, \epsilon) \gamma_{\mathbf{n},p}^{S'}(\mathbf{u}, \epsilon) \, d\mathbf{u}$

Use the translation invariance of the system to derive the spatio-spectral covariance matrix for a homogeneously illuminated detector: $\mathrm{COV}_\epsilon^{S,S'}(\mathbf{k}) = \sum_\mathbf{m} \mathrm{cov}_\epsilon(y_\mathbf{m}^S, y_{\mathbf{m}+\mathbf{k}}^{S'})$.

(7) $\quad \mathrm{COV}_\epsilon^{S,S'}(\mathbf{k}) = \sum_\mathbf{m} \mathrm{cov}_\epsilon(y_\mathbf{m}^S, y_{\mathbf{m}+\mathbf{k}}^{S'})$.

- Calculate the Optical Transfer Function $\mathrm{OTF}^S(\mathbf{v}, \epsilon)$ by a 2D Fourier transform of the $\mathrm{PSF}^S(\mathbf{v}, \epsilon)$ and the Noise Power spectrum $\mathrm{NPS}^{S,S'}(\mathbf{v}, \epsilon)$ by a discrete 2D Fourier transform of $\mathrm{COV}_\epsilon^{S,S'}(\mathbf{k})$.
- For a given spectrum $\phi(\epsilon)$, calculate a weighted average of OTF and NPS:

(8) $\quad \mathrm{OTF}^S(\mathbf{v}) = \int \phi(\epsilon) \mathrm{OTF}^S(\mathbf{v}, \epsilon) \, d\epsilon$,

(9) $\quad \mathrm{NPS}^{S,S'}(\mathbf{v}) = \int \phi(\epsilon) \mathrm{NPS}^{S,S'}(\mathbf{v}, \epsilon) \, d\epsilon$,

where $\phi(\epsilon)$ is the normalized spectrum: $\int \phi(\epsilon) \, d\epsilon = 1$.

A quantity that helps interpreting the results is the total response $R^S(\epsilon)$, the number of counts created from a single photon of energy $\varepsilon$ in all pixels in some spectral channel $s$. Due to of double counting, the total response can be greater than 1. The spectral response is simply the OTF at zero spatial frequency:

(10) $\quad R^S(\epsilon) = \iint \mathrm{PSF}^S(\mathbf{v}, \epsilon) \, d\mathbf{v} = \mathrm{OTF}^S(\mathbf{v} = 0, \epsilon)$.

### 2.3 Calculation of spectral DQEs

From the OTF and the diagonal elements of the NPS, it is straightforward to calculate the Detective Quantum Efficiency (DQE) for an individual spectral channel $s$:

(11) $\quad \mathrm{DQE}^S(\mathbf{v}, \epsilon) = \frac{(\mathrm{OTF}^S(\mathbf{v},\epsilon))^2}{\mathrm{NPS}^{S,S}(\mathbf{v},\epsilon)}$.

This is the classical DQE describing the signal/noise performance of each spectral channel $s$ separately as a function of spatial frequency and photon energy. It is obvious, however, that this DQE is useless when it comes to quantifying the *spectral* performance of a detector with several spectral channels. To obtain useful DQE-like quantities relevant for *spectral* detectors, it is necessary to specify spectral tasks related to some typical clinical tasks. We follow closely the approach of Persson et al. [4], Tanguay et al. [5] and Zarif Yussefian and Tanguay [6]. Note that the DQEs developed there make statements regarding the *information* contained in the spectral channels, in the sense of a Cramér-Rao lower bound for the variance of any unbiased estimator. There is no guarantee that there is an estimator that can efficiently use this information.

We now show how the spectral DQEs can be calculated from the quantities prepared in the previous section. First, we can obtain the task-independent threshold response from the OTF and the normalized spectrum $\phi(\epsilon)$:

(12) $\quad T^S(\mathbf{v}) = \int \phi(\epsilon) \mathrm{OTF}^S(\mathbf{v}, \epsilon) \, d\epsilon$,

the total response of spectral channel $s$ is given by

(13) $\quad c_0^S = T^S(0) = \int \phi(\epsilon) \mathrm{OTF}^S(0, \epsilon) \, d\epsilon = \int \phi(\epsilon) R^S(\epsilon) \, d\epsilon$,

the normalized spectral noise power spectrum can be determined as

$$\text{(14)} \quad \text{NNPS}^{s,s\prime}(\nu) = \frac{\text{NPS}^{s,s\prime}(\nu)}{c_0^s c_0^{s\prime}}.$$

Denoting the materials with the index $b$, we determine the normalized response of spectral channel $s$ to an attenuation from material $b$

$$\text{(15)} \quad M_{s,b}(\nu) = \frac{\int \phi(\epsilon)\left(\frac{\mu}{\rho}\right)_b \text{OTF}^s(\nu,\epsilon) d\epsilon}{c_0^s}.$$

The matrix of 2nd moments $\mathbf{M}_{b,b\prime}$ can be expressed as

$$\text{(16)} \quad \mathbf{M}_{b,b\prime} = \int \phi(\epsilon) \left(\frac{\mu}{\rho}\right)_b \left(\frac{\mu}{\rho}\right)_{b\prime} d\epsilon.$$

Putting things together, the task-independent DQE can be written as

$$\text{(17)} \quad \text{DQE}^{\text{task-ind}}(\nu) = \sum_{s,s\prime} T^s(\nu)(\text{NPS}^{-1}(\nu))_{s,s\prime} T^{s\prime}(\nu),$$

the DQE for the spectral task of material *decomposition* is given by

$$\text{(18)} \quad \text{DQE}_b^{\text{decomp}}(\nu) = \frac{(\mathbf{M}^{-1})_{b,b}}{(\mathbf{Q}^{-1})_{b,b}},$$

where $\mathbf{Q}$ is the frequency-dependent matrix

$$\text{(19)} \quad \mathbf{Q}_{b,b\prime}(\nu) = \sum_{s,s\prime} M_{s,b}(\nu)(\text{NNPS}(\nu))^{-1}_{s,s\prime} M_{s\prime,b\prime}(\nu).$$

The DQE for the non-spectral task of *quantification* of a *known material b* (called *detection* in [6]) depends only on the diagonal elements of the $\mathbf{Q}$ and $\mathbf{M}$ matrices:

$$\text{(20)} \quad \text{DQE}_b^{\text{quant}}(\nu) = \frac{\mathbf{Q}_{b,b}(\nu)}{\mathbf{M}_{b,b}}.$$

Note that, while this is a non-spectral task because the material is known, it will make use of all spectral channels in an optimal way. In the following, we will investigate material decomposition and quantification tasks for the clinically materials water and iodine.

## 2.4 Detectors considered

Four types of realistic photon counting detectors were analyzed:

   a. CdTe based direct converting photon counting detector,
   b. Si based direct converting photon counting detector,
   c. CdTe based photon counting detector with additional coincidence counters,
   d. LaBr$_3$ based optical counting detector.

Both CdTe and Si based direct converting detectors have already been investigated in some detail (e.g. [1], [3]-[6]). New contributions in this paper are assessments of the spectral performance of a photon counter with coincidence counter and an optical counting detector.

All x-ray Monte Carlo simulations were performed with MOCASSIM (Siemens Healthcare).

## 2.5 Modelling CdTe

The simulation of the CdTe photon counting detector is done in exactly the same way as in [3]. The geometry is sketched in Figure 2.

For CdTe, the pixel structure is generated by the back layer electrodes only.

For the CdTe simulations, a 5x5 pixel area was considered. The PSF was calculated subdividing the pixel into 100x100 subpixels, simulating 1000 photon histories for each subpixel.

## 2.6 Modelling the Si detector

The geometry of the silicon detector is sketched in Figure 3 (not to scale). The Monte Carlo simulation follows Persson et al. [9]. In contrast to the CdTe detector where the absorbing semiconductor can be considered as isotropic, the tungsten foils are breaking the symmetry. This means that the Monte Carlo simulation for the x-ray interactions has to be done for several impact locations relative to the foils. We chose 10 different locations along the direction perpendicular to the foils. The lower absorption of Silicon calls for an edge-on design where the incoming photons impinge on the edge of a Si wafer and electrodes run along the depth of the material. To reduce Compton scatter, 20 µm tungsten foils (shown in black) are separating the pixels in one dimension. The electrodes can be subdivided along the depth dimension which reduces the number of counts seen in one electrode, mitigating the pileup issue. This has negligible effect in the low count rate limit, however, and has not been used in the simulation.

Due to the large mean free path of x-ray photons in silicon, scattered photons can travel far in this material and the simulation needs to include a relatively large area. We chose to simulate the silicon detector using an area of 13x13 pixels, subdividing the pixel into a non-isotropic grid of 10 subpixels in x (perpendicular to the direction of the tungsten foils) and 500 subpixels in y (parallel to the foils). The high oversampling in y is needed to see charge sharing events in Si where the charge cloud width is smaller and the pixel size larger than with CdTe. But still charge sharing is an almost negligible process in Si. Due to the tungsten foils, there is no charge sharing in x direction, so crosstalk is mainly related to Compton scattered x-ray photons and a fine sampling is unnecessary. For each subpixel, 100 photon histories were used.

The tungsten foils introduce anisotropy into the silicon system that in principle affects all quantities. In reality, however, this anisotropy is very small. We will show results only for the direction perpendicular to the tungsten foils.

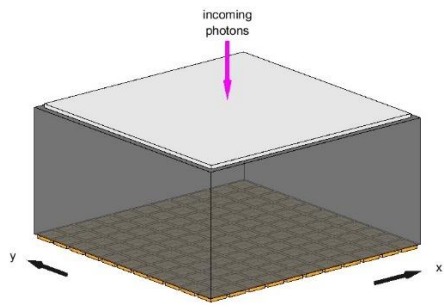

*Figure 2: Geometry of the CdTe detector. The pixel structure is created by the structured electrodes (shown in yellow) at the back side.*

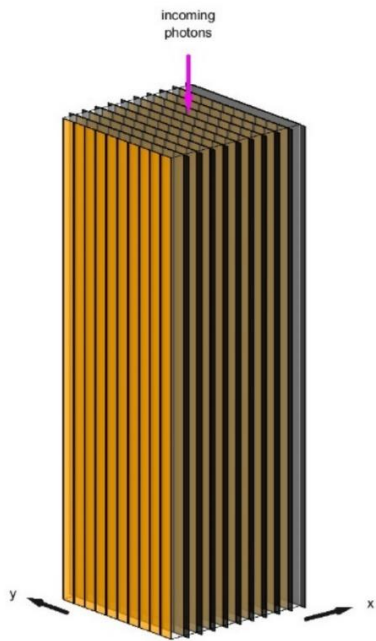

*Figure 3: Geometry of the Si detector. The pixel structure is given by the thickness of the Si wafers placed edge-on towards the incoming photons and the strip electrodes on the side of the wafers. The separating tungsten foils are shown in black. Not to scale!*

The basic parameters of the various detector materials are listed in Table 1:

|  | Density (g/cm$^3$) | Thickness (mm) | Pixel size (mm) | Electronic noise (keV) | Charge cloud width (µm) | Remarks |
|---|---|---|---|---|---|---|
| CdTe [3] | 5.85 | 1.6 | 0.3 | 1.5 | 23 |  |
| Si [4] | 2.33 | 60 | 0.5 | 1.5 | $1.9 \cdot \left(\frac{E}{1\ keV}\right)^{0.54}$ | 20 µm W foils |
| LaBr$_3$ [9] | 5.1 | 3 | 0.5 | n.a. | n.a. | Simplifying assumptions see section 2.7 |

*Table 1: Basic parameters for the direct converting detectors*

### 2.7 CdTe photon counting detector with additional coincidence counters

Simulations for this detector are based on the simulations of the CdTe detector without coincidence counters described in section 2.5 and [3]. The detector was modeled with 2 counters for energy thresholds 20 and 70 keV and 3 additional coincidence counters. Coincidence counters are additional counter which trigger when the charge deposited in one pixel exceeds a certain energy threshold and one of the neighboring pixels exceeds another energy threshold at the same time. For low flux such counters count only charge sharing / fluorescence events and can be used to greatly increase the performance of a photon counting detector and have been studied extensively [7],[8].

In this study, three such counters were considered. The first counter is triggered when both the charge in the pixel itself, as well as in any of its eight direct neighbors exceeds the lowest threshold of 20 keV. The second counter is triggered when the pixel under consideration exceeds the upper threshold of 70 keV and any neighbor exceeds the lower threshold. The third counter is triggered when the pixel under consideration exceeds the lower and any neighbor exceeds the higher threshold. No coincidence counter for events triggering the higher threshold in two pixels was simulated, as such events are very, very unlikely for a 140 kV x-ray spectrum.

### 2.8 Optical photon counting detector

The simulation of the optical counting detector is based on the following simplifying assumptions [9]:

- The x-rays interact with the scintillator; the conversion of the absorbed energy into optical photons follows a Poisson distribution (assumption: 50 optical photons per absorbed keV, Fano factor 1).
- All optical photons travel randomly to the bottom of the pixel to one of 625 Si-photomultipliers (PM). For a PM hit by *n* optical photons, the probability of firing is assumed to be $1 - (1 - PDE)^n$. We assume a photo detection efficiency *PDE* = 0.3.
- The quantity that will be compared to the thresholds is the number of PMs firing within a short time interval. Naturally, this quantity saturates at 625, the number of PMs. This saturation creates a highly non-linear *spectral* response. But note that the detector is still a *linear* detector in the system theoretical sense: as long as we stay in the low flux limit, doubling the flux will double the signal.

The whole process is simulated in an end-to-end Monte Carlo simulation containing all four random steps (x-ray absorption, optical photon production, spread to the PMs, firing of the PMs).

The signal generation process is illustrated in Figure 4.

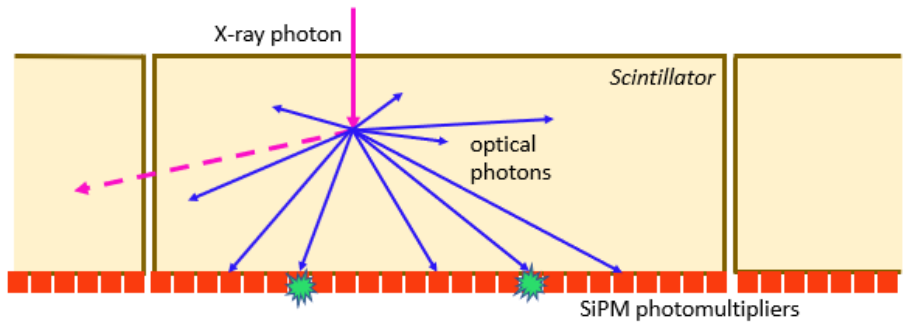

*Figure 4: Schematics of the signal generation in the optical detector system. A count is registered if the number of SiPMs firing exceeds a given threshold.*

For optical counting, there is no charge sharing effect and crosstalk is mainly related to K fluorescence and Compton scatter. The septa between the pixels are assumed to be perfect reflectors and their width is neglected. We subdivide the pixel into 25x25 subpixels, simulating 2000 photon histories per subpixel. Effects like optical crosstalk due to secondary triggering of SiPMs [11] have been neglected.

## 3. RESULTS

For the evaluation of the various DQEs, we assume a spectrum clinically relevant for CT applications of 140 kV, filtered by 30 cm of water (cf. Figure 5).

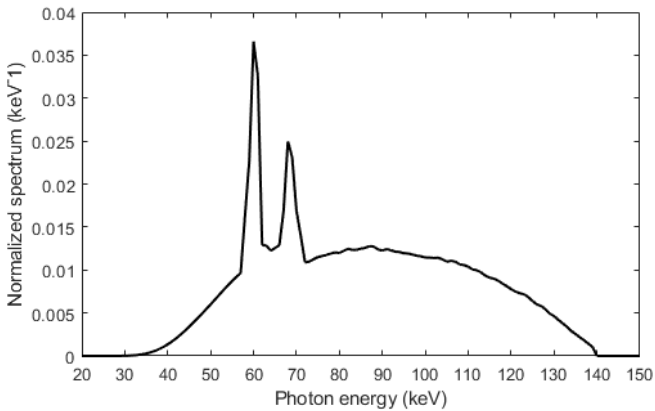

*Figure 5: X-ray spectrum used for the calculation of the spectral DQEs: 140 kV, filtered with 30 cm of water.*

Results of the calculations for the four detector systems are shown in the following figures.

To give an impression of the spectral properties of the different detectors, we plot the energies detected from an 80 keV photon (Figure 6). This kind of plot is generated by sweeping the energy threshold across the relevant range and detecting the signals detected. Differentiating with respect to the threshold gives the spectra shown. Note that this kind of plot does not make sense for systems with fixed thresholds like the CdTe detector with coincidence counters. As expected, the peak is located at the input photon energy, somewhat broadened due to electronic noise (CdTe, Si) or other statistical effects (LaBr$_3$). For lower output energies, CdTe shows two peaks at the typical K fluorescence energy (~24 keV) and the complementary energy ((80 – 24) keV = 56 keV) plus a substantial, almost flat, spectrum due to charge sharing. The Si spectrum looks much cleaner, with a narrow photo peak and a Compton plateau starting around 20 keV. However, not many photons end up in the photon peak which makes it necessary to use also very low thresholds below 10 keV which carry some, but limited, spectral information [12]. A not very relevant but interesting feature of the Si plot is the little bump at 59 keV which stems from the photons that underwent a K fluorescence from the tungsten foils. The optical counting system has a very broad photo peak that contains most of the photons. There is no charge sharing, so the only feature in the plot below the photo peak are the K fluorescence peak at ~33 keV and its complementary peak at ~47 keV.

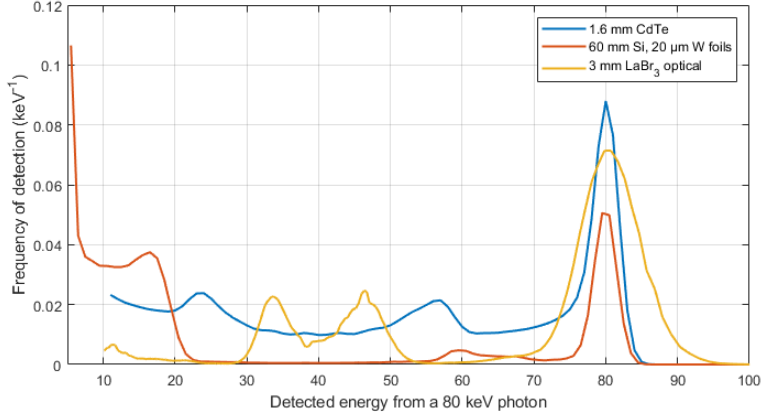

Figure 6: Density of energies detected from a single incoming 80 keV photon. Typical spectra (see Figure 5) have no photon energies below 30 keV, so there is no need to consider very low thresholds. This is different for the Si detector where very low thresholds are needed to detect photons that are only Compton scattered.

Figure 7 shows the spectral responses, that is, the number of counts generated in any pixel from an incoming photon as a function of photon energy. Figure 8 shows the various spectral DQEs defined by the formula in section 2.3.

### 4.1 Spectral response

For all considered detector systems, Figure 7 shows the spectral responses for all spectral channels used. The spectral response can be larger than 1 if double counts in several neighboring pixels are generated. This is most prominently the case in CdTe (mainly due to charge sharing and K-fluorescence) and in LaBr$_3$ (due to K-fluorescence).

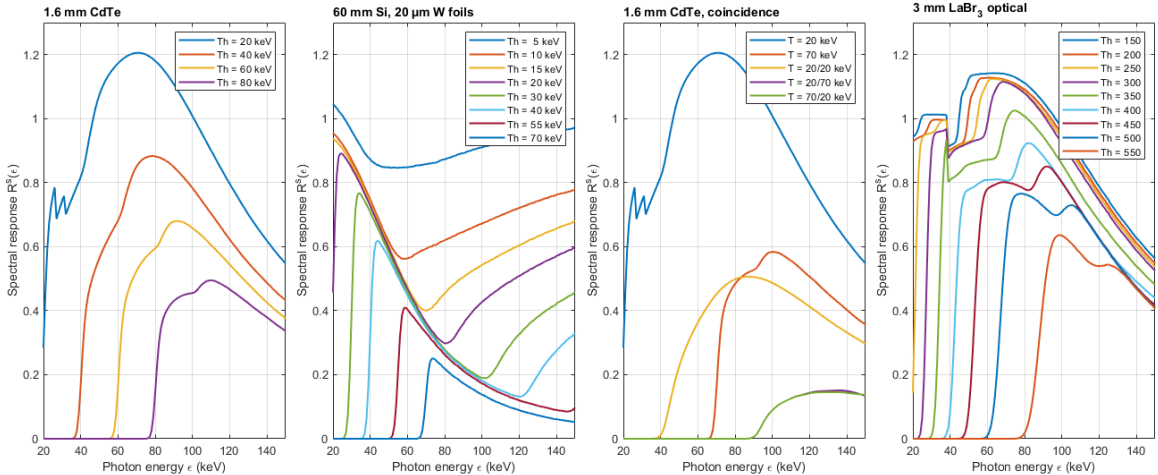

Figure 7: Spectral responses $R^s(\epsilon)$ of the four different detector systems for the spectral channels used. Plotted is the average count signal in the spectral channel, summed over all pixels, for an incoming photon of a given energy.

The curves for CdTe show the spectral distortion due to charge sharing: For an ideal detector, a photon of 100 keV should on the average create the same signal in all thresholds below 100 keV. The curves show, however, that the signal for the lower thresholds is substantially higher than for the 80 keV threshold.

The plots for Si show a quite different behavior: the spectral responses show a drop as a function of photon energy, followed by an increasing response for photon energies. The drop is explained by the usual drop of the photo absorption cross section with energy; the signal starts to rise again when the energy loss due to Compton scatter exceeds the threshold. This creates a somewhat paradox situation where the response at higher energies relies mainly on the availability of low thresholds.

The spectral response of the CdTe system with two thresholds (20 and 70 keV) and three coincidence counters is easy to understand: The two threshold responses behave analogous to thresholds of the CdTe detector, the coincidence counters can be triggered only when the photon energy exceeds the sum of the two involved thresholds.

The optical counting system behaves similar as the CdTe system. Some double counting can be observed, but the spectral contamination is much less pronounced than with CdTe: Beyond the various thresholds, the curves for the spectral responses of the thresholds are much closer.

### 4.2 Spectral DQE

The spectral Detective Quantum Efficiencies according to equations (17), (18) and (20) for the x-ray spectrum of Figure 5 and the materials water and iodine are shown in Figure 8. For the Si detector, the direction orthogonal to the tungsten foils is shown.

For standard imaging without any spectral component, the solid blue curve (water contrast) is most relevant. Here, the good x-ray absorption across the spectrum favors the CdTe and $LaBr_3$ systems.

In all cases, the DQEs for true spectral tasks (broken lines) are substantially lower than the DQEs for the water or iodine quantification tasks (solid lines): true spectral decomposition is a much harder task than just estimating the thickness of a known material. For these spectral tasks and for spatial frequencies less than 0.7 $mm^{-1}$, Si has an advantage over CdTe, but both are beaten by the optical counting system which is only handicapped by its relatively large 0.5 mm pixels.

A CdTe detector with additional coincidence channels can, at least in a Cramér-Rao lower bound sense, restore much of the spectral performance lost due to charge sharing. Only the optical counting system shows better spectral performance than this detector. Coincidence channels can also substantially improve the non-spectral performance, simply because they provide information about double counting events.

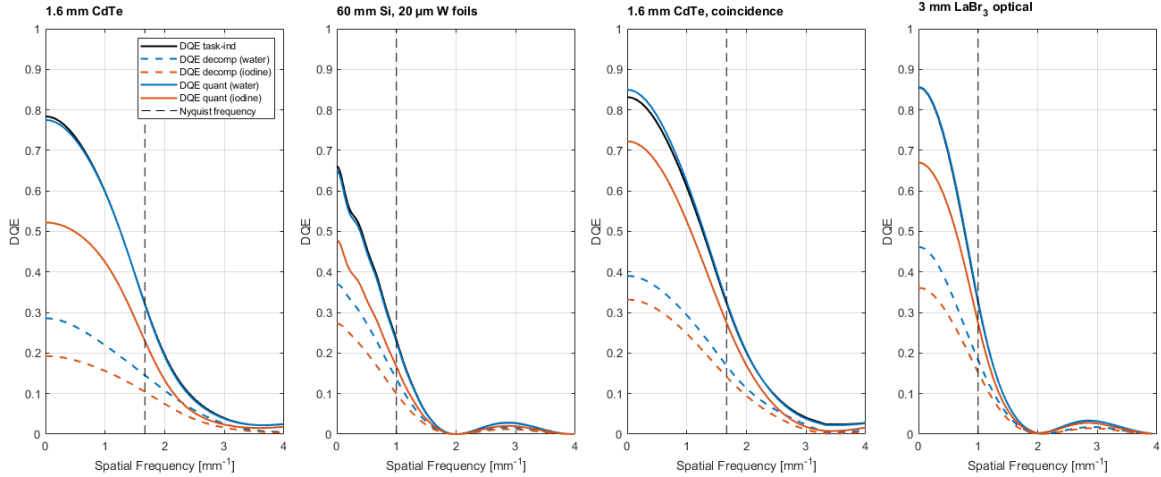

*Figure 8: Spectral and non-spectral DQEs according to equations (17), (18) and (20) (same spectral channels as in Figure 7) for the four detector types considered. The black lines correspond to the task-independent spectral DQE of equation (17). The dashed lines show the DQEs for water and iodine in a water/iodine decomposition task, the colored solid lines correspond to the non-spectral tasks of estimating the water and iodine thickness.*

## 4. SUMMARY

We have presented a method to obtain all relevant performance characteristics for spectral pixelized detectors from an end-to-end Monte Carlo simulation.

The method was demonstrated for four different realistic detectors: Direct converting detectors using CdTe and Silicon as semiconductors, a CdTe detector with additional coincidence counters and an optical counting system. To our knowledge, the two latter systems have not been previously investigated in this depth.

Out of the two realized detectors (CdTe and Si), the CdTe system has very good standard performance (water or iodine quantification) up to high spatial frequencies. Silicon has an advantage when it comes to true spectral performance at lower spatial frequencies. Both detectors are beaten only by the so far unrealized coincidence and optical counting systems.

The limitations of this study are given by the simplifying assumptions: In no case did we consider any effects of higher count rates (e.g. pulse pileup). Particularly for the optical counting system, we made several assumptions like ideally reflecting, x-ray transparent but infinitely thin septa between the pixels or identical sensitivity for all SiPMs. For CdTe, the models in the low-count regime have already been cross-validated with measurements (cf. [3]), but the performance assessment of CdTe with coincidence counters based on the spectral DQEs can be exploited only if estimators can be found that come close to these bounds. The same is true for the Si detector where it might be a non-trivial problem to extract the information contained in the low keV thresholds which may also be sensitive to electronic noise.

An unexpected result of our investigation was the good performance of the $LaBr_3$-based optical counting system. If the high-flux issue could be solved and if it was possible to design optical separators that allow using smaller pixel sizes without sacrificing too much active area, such systems might in future become an interesting alternative to direct converting semiconductors.